\documentclass[preprintnumbers,prd,showpacs,floatfix,superscriptaddress,nofootinbib,twocolumn]{revtex4-1}
\usepackage{graphicx,epsfig}
\usepackage{amsmath}
\usepackage {amssymb}
\usepackage {longtable}
\usepackage{multirow}
\usepackage{dcolumn}
\usepackage{bm}
\usepackage{amsfonts}
\usepackage{subfigure}
\usepackage{color}
\usepackage{relsize}

\begin{document}
%%%%%%%%%%%%%%%%%%%%%%%%%%%%%%%%%%%%%%%%%%%%%%%%%%%%%%%%%%%%%%%%%%%%

%%%%%%%%%%%%%%%%%%%%%%%%%%%%%%%%%%%%%%%%%%%%%%%%%%%%%%%%%%%%%%%%%%%%
\title{Thick brane structures in generalized hybrid metric-Palatini gravity}
%-----------------------------------------------------------------
\author{Jo\~{a}o Lu\'{i}s Rosa}
\email{joaoluis92@gmail.com}
\affiliation{Institute of Physics, University of Tartu, W. Ostwaldi 1, 50411 Tartu, Estonia}
%-----------------------------------------------------------------
\author{Douglas A. Ferreira}
\email{ferreira.douglasdaf@gmail.com}
\affiliation{Departamento de F\'isica, Universidade Federal da Para\'iba, 58051-970, Jo\~ao Pessoa, PB, Brazil}
%-----------------------------------------------------------------
\author{Dionisio Bazeia}
\email{dbazeia@gmail.com}
\affiliation{Departamento de F\'isica, Universidade Federal da Para\'iba, 58051-970, Jo\~ao Pessoa, PB, Brazil}
%-----------------------------------------------------------------
\author{Francisco S. N. Lobo}
\email{fslobo@fc.ul.pt}
\affiliation{Instituto de Astrofísica e Ci\^encias do Espa\c{c}o, Faculdade de Ci\^encias da Universidade de Lisboa, Edif\'icio C8, Campo Grande, P-1749-016, Lisbon, Portugal}
%-----------------------------------------------------------------
%%%%%%%%%%%%%%%%%%%%%%%%%%%%%%%%%%%%%%%%%%%%%%%%%%%%%%%%%%%%%%%%%%%%

%%%%%%%%%%%%%%%%%%%%%%%%%%%%%%%%%%%%%%%%%%%%%%%%%%%%%%%%%%%%%%%%%%%%
\date{\today}
%%%%%%%%%%%%%%%%%%%%%%%%%%%%%%%%%%%%%%%%%%%%%%%%%%%%%%%%%%%%%%%%%%%%

%%%%%%%%%%%%%%%%%%%%%%%%%%%%%%%%%%%%%%%%%%%%%%%%%%%%%%%%%%%%%%%%%%%%
\begin{abstract} 
In this work, we study 5-dimensional braneworld scenarios in the scalar-tensor representation of the generalized hybrid metric-Palatini gravitational theory. We start by considering a model for a brane supported purely by the gravitational scalar fields of the theory and then consider other distinct cases where the models are also supported by an additional matter scalar field. We investigate the stability of the gravity sector and show that the models are all robust against small fluctuations of the metric. In particular, in the presence of the additional scalar field, we find that the profile of the gravitational zero mode may be controlled by the parameters of the model, being also capable of developing internal structure.   
\end{abstract}
%%%%%%%%%%%%%%%%%%%%%%%%%%%%%%%%%%%%%%%%%%%%%%%%%%%%%%%%%%%%%%%%%%%%

%%%%%%%%%%%%%%%%%%%%%%%%%%%%%%%%%%%%%%%%%%%%%%%%%%%%%%%%%%%%%%%%%%%%
%%%%%%%%%%%%%%%%%%%%%%%%%%%%%%%%%%%%%%%%%%%%%%%%%%%%%%%%%%%%%%%%%%%%

%%%%%%%%%%%%%%%%%%%%%%%%%%%%%%%%%%%%%%%%%%%%%%%%%%%%%%%%%%%%%%%%%%%%
\maketitle
%%%%%%%%%%%%%%%%%%%%%%%%%%%%%%%%%%%%%%%%%%%%%%%%%%%%%%%%%%%%%%%%%%%%

%%%%%%%%%%%%%%%%%%%%%%%%%%%%%%%%%%%%%%%%%%%%%%%%%%%%%%%%%%%%%%%%%%%%
\section{Introduction}\label{sec:intro}
%%%%%%%%%%%%%%%%%%%%%%%%%%%%%%%%%%%%%%%%%%%%%%%%%%%%%%%%%%%%%%%%%%%%

Since its development in 1915, General Relativity (GR) has been a very successful theory, at least for local tests. However, the discovery of the late-time cosmic acceleration \cite{Perlmutter:1998np,Riess:1998cb} has spurred research in modified dynamics at large scales. In this context, the possibility that the Hilbert-Einstein term is supplemented with more general combinations of curvature invariants has been extensively explored \cite{fRgravity8, Capozziello:2002rd,Capozziello:2011et, DeFelice:2010aj, Lobo:2008sg, fRgravity7, Nojiri:2010wj,rev2}. Indeed, it was soon found that the usual metric formulation, which considers that the metric is the fundamental field, differs generically from its Palatini (or metric-affine) counterpart \cite{Olmo}, where here the metric and the connection are assumed to be the two fundamental fields of the theory. The metric approach leads to higher-order derivative equations, contrary to the field equations in the Palatini formulation, which are second-order. 
We do emphasize that the purely $f(R)$ metric and metric-affine formalisms coincide in GR, however, the two formalisms lead to different results considering more generic functions $f(R)$ \cite{Olmo}.
For instance, the scalar-tensor representation of $f(R)$ gravity is useful to illustrate the differences between the metric and Palatini approaches. In the metric formalism, the scalar field behaves as a dynamical field, which satisfies a modified Klein-Gordon equation with self-interactions that essentially depend on the form of $f(R)$. However, the scalar field $\phi$ should have a very low mass, which implies a long interaction range, in order to have an impact at large scales; nevertheless, light scalars do indeed have an impact at smaller scales, and are strongly constrained by local observations, at the laboratory and Solar System scales, unless screening mechanisms are invoked \cite{screen4,screen2,screen5}. In the scalar-tensor representation of the Palatini formalism, the scalar field satisfies an algebraic function of the trace of the matter stress-energy tensor, which lead to undesired gradient instabilities \cite{Koivisto:2006ie,Koivisto:2005yc,Olmo:2006zu,Olmo:2008ye}.

However, these difficulties may be avoided within a {\it hybrid} variation of these theories, in which the purely metric Einstein-Hilbert action is supplemented with a metric-affine correction term \cite{Capozziello:2013uya,Harko:2011nh,Capozziello:2012ny,Capozziello:2015lza,Harko:2020ibn}. More specifically, an interesting aspect of these theories is the possibility of generating long-range forces without conflicts with the local tests and without invoking screening mechanisms. 
The possibility of expressing these hybrid $f(R)$ metric-Palatini theories using a scalar-tensor representation simplifies the analysis of the field equations and the construction of solutions. 
It is interesting to note that this theory, which in the linear approach takes the form $R+f({\cal R})$, one retains the positive results through the Einstein-Hilbert term $R$ and the additional gravitational corrections are given by the metric-affine $f({\cal R})$ component, where the Palatini curvature scalar ${\cal R}$ is constructed in terms of an independent connection. 
A wide variety of applications of this hybrid metric-Palatini theory has been explored, namely, in considering that dark matter is a geometric effect of modified gravity \cite{Capozziello:2012qt,Capozziello:2013yha}, in exploring the Cauchy problem \cite{Capozziello:2013gza} and the Noether symmetries \cite{Borowiec:2014wva}, black hole, wormhole and stellar solutions \cite{Capozziello:2012hr,Danila:2018xya, Bronnikov:2019ugl,Bronnikov:2020vgg,Danila:2016lqx}, the Einstein static Universe \cite{Boehmer:2013oxa}, string-like configurations \cite{Harko:2020oxq,Bronnikov:2020zob}, and thick branes \cite{Fu:2016szo}, among others.

The linear hybrid metric-Palatini theory can be further generalized, where the gravitational action depends on a general function of both the metric and Palatini curvature scalars \cite{Tamanini:2013ltp}. This extension has also received attention with a plethora of applications, namely, cosmological solutions \cite{Rosa:2017jld,Rosa:2019ejh}, weak-field phenomenology \cite{bombacigno}, wormhole \cite{Rosa:2018jwp} and black hole \cite{Rosa:2020uoi} solutions. In this work, we consider the possibility to study braneworld structures in the generalized hybrid metric-Palatini gravity. In this context, as is well-known, the Randall-Sundrum braneworld model originally proposed \cite{RS} was soon generalized \cite{GW,Freedman,Csaki} to describe thick braneworld scenarios in the presence of scalar fields, which are included to source the warped 5-dimensional AdS$_5$ geometry with a single extra spatial dimension of infinite extent. Soon after, several authors investigated the braneworld scenario with a diversity of motivations; see, e.g., Refs. \cite{gremm,brito,lisa,brito2,gomes,Bazeia:2004dh,deSouzaDutra:2008gm,andri,cuba,B1,B2,B3,B4,almeida,almeida2,dutra,correia,liu1,liu2,liu3,danilo,douglas,riazi,roldao,rocha,ahmed} and references therein for many distinct possibilities to implement investigations on thick braneworlds, including the absence of scalar fields, the presence of two scalar fields with standard dynamics, the case of tachyonic and other generalized dynamics, asymmetric thick brane, and also the $f(R)$, Gauss-Bonnet and several other possibilities of modifications of GR \cite{DeFelice:2010aj,Nojiri:2010wj,Capozziello:2011et,rev2}.
  
In order to deal with the 5-dimensional generalized hybrid metric-Palatini braneworld scenario, the study developed in the present work is outlined in the following manner: in Sec. \ref{sec:equations} we introduce the model and write the equations of motion. Furthermore, in Sec. \ref{sec:models} we consider the warped geometry with a single extra dimension of infinite extent, rewrite the equations of motion on general grounds and show how to get to the thick braneworld scenarios in the case of standard GR. We also describe distinct solutions of current interest, in the absence and in the presence of an extra field $\chi$. In Sec. \ref{sec:perturb}, we discuss the robustness of the geometric sector in the generalized hybrid metric-Palatini braneworld theory. Finally, in Sec. \ref{sec:concl}, we summarize our results and conclude.
Throughout this work, we implement all the calculations considering a system of units for which $c=1$, using capital latin indexes $\{M,N,...\}$ running from 0 to 4 and greek indexes $\{\mu,\nu,...\}$ running from 0 to 3.

%%%%%%%%%%%%%%%%%%%%%%%%%%%%%%%%%%%%%%%%%%%%%%%%%%%%%%%%%%%%%%%%%%%%
\section{Action and field equations}\label{sec:equations}
%%%%%%%%%%%%%%%%%%%%%%%%%%%%%%%%%%%%%%%%%%%%%%%%%%%%%%%%%%%%%%%%%%%%

The generalized hybrid metric-Palatini theory \cite{Tamanini:2013ltp,Rosa:2017jld} in 4+1 dimensional gravity  is described by an action functional $S$ of the form 
\begin{equation}\label{act1}
S=\frac{1}{2\kappa^2}\int_\Omega\sqrt{-g}f\left(R,\cal{R}\right)d^5x+S_m\left(g_{MN},\chi\right),
\end{equation}
where $\kappa^2\equiv 8\pi G_5$, $G_5$ is the 5-dimensional Newtonian constant, $\Omega$ is a 5-dimensional spacetime manifold on which we define the coordinate set $x^M$, $g$ is the determinant of the metric $g_{MN}$, $f$ is an arbitrary function of the Ricci scalar, $R=g^{MN}R_{MN}$, where $R_{MN}$ is the Ricci tensor, and the Palatini scalar curvature $\mathcal R=g^{MN}\mathcal R_{MN}$, where the Palatini Ricci tensor $\mathcal R_{MN}$ is defined in terms of an independent connection $\hat\Gamma^P_{MN}$ as
\begin{equation}
\mathcal{R}_{MN}=\partial_P
\hat\Gamma^P_{MN}-\partial_N\hat\Gamma^P_{MP}+\hat\Gamma^P_{PQ}\hat\Gamma^Q_{MN}-\hat\Gamma^P_{MQ}\hat\Gamma^Q_{PN}.
\end{equation}
$S_m$ is the matter action defined as $S_m=\int d^5x\sqrt{-g}\;{\cal L}_m$ where ${\cal L}_m$ is the matter Lagrangian density considered minimally coupled to the metric $g_{MN}$, and $\chi$ collectively denotes the matter fields. 

Taking the variation of Eq. \eqref{act1} with respect to the independent connection $\hat\Gamma^P_{MN}$ yields the equation of motion
\begin{equation}\label{varcon}
\hat\nabla_P\left(\sqrt{-g}\frac{\partial f}{\partial \cal{R}}g^{MN}\right)=0,
\end{equation}
where $\hat\nabla$ denotes the covariant derivative written in terms of the independent connection $\hat\Gamma^P_{MN}$. Recalling that $\sqrt{-g}$ represents a scalar density of weight 1, we have that $\hat\nabla_P \sqrt{-g}=0$, and thus Eq. \eqref{varcon} implies the existence of a new metric tensor $h_{MN}=\left(\partial f/\partial\mathcal R\right) g_{MN} $, conformally related to the metric $g_{MN}$ with a conformal factor $\partial f/\partial\mathcal R$, for which the connection $\hat\Gamma^P_{MN}$ is the Levi-Civita connection, i.e., one can write $\hat\Gamma^P_{MN}$ as
\begin{equation}
\hat\Gamma^P_{MN}=\frac{1}{2}h^{PQ}\left(\partial_M h_{QN}+\partial_N h_{MQ}-\partial_Q h_{MN}\right),
\end{equation}
where $\partial_M$ denotes partial derivatives. The conformal relation between the metrics $h_{MN}$ and $g_{MN}$ implies that the two Ricci scalars $R_{MN}$ and $\mathcal R_{MN}$, assumed \textit{a priori} as independent, are in fact related by the expression
\begin{eqnarray}\label{ricrel}
\mathcal R_{MN}=R_{MN}-\frac{1}{f_\mathcal R}\left(\nabla_M\nabla_N+\frac{1}{3}g_{MN}\Box\right)f_\mathcal R
	\nonumber  \\
+\frac{4}{3f_\mathcal R^2}\partial_M f_\mathcal R\partial_N f_\mathcal R,
\end{eqnarray}
where $\nabla_M$ denotes covariant derivatives written in terms of the connection $\Gamma^P_{MN}$, $\Box=\nabla^P\nabla_P$ is the D'Alembert operator, and the subscripted $f_\mathcal R$ denote derivatives of $f$ with respect to $\mathcal R$. 

An equivalent scalar-tensor representation of Eq. \eqref{act1} can be obtained via the definition of two scalar fields $\varphi$ and $\psi$ and a scalar potential $V\left(\varphi, \psi\right)$ as
\begin{equation}\label{scalars}
\varphi=\frac{\partial f}{\partial R},\qquad \psi=-\frac{\partial f}{\partial\mathcal R}\,,
\end{equation}
\begin{equation}\label{potential}
V\left(\varphi,\psi\right)=\varphi R -\psi\mathcal R-f\left(R,\mathcal R\right).
\end{equation}
Inserting both Eqs. \eqref{scalars} and \eqref{potential} in Eq. \eqref{act1}, using the trace of Eq. \eqref{ricrel} to cancel the factor $\mathcal R$, writing $f_\mathcal R=-\psi$ and neglecting the boundary term proportional to $\Box\psi$, we arrive to the scalar-tensor representation of the generalized hybrid metric-Palatini gravity as
\begin{eqnarray}
S=\frac{1}{2\kappa^2}\int_\Omega \sqrt{-g}&&\left[\left(\varphi-\psi\right) R-\frac{4}{3\psi}\partial^P\psi\partial_P\psi 
	\right.
		\nonumber \\
&&\left.-V\left(\varphi,\psi\right)\right]d^5x+S_m\left(g_{MN},\chi\right).\label{act2}
\end{eqnarray}

Without further consideration of the matter fields $\chi$, there are three independent variables in Eq. \eqref{act2}, these are the metric $g_{MN}$ and the two scalar fields $\varphi$ and $\psi$. Varying Eq. \eqref{act2} with respect to the metric $g_{MN}$ yields the modified field equations
\begin{eqnarray}
\left(\varphi-\psi\right)G_{MN}+\frac{1}{2}g_{MN}\left[\frac{4}{3\psi}\partial^P\psi\partial_P\psi+V\left(\varphi,\psi\right)\right]\label{eqfield}
	\nonumber  \\
-\frac{4}{3\psi}\partial_M\psi\partial_N\psi-\left(\nabla_M\nabla_N-g_{MN}\Box\right)\left(\varphi-\psi\right)
	\nonumber  \\
=\kappa^2T_{MN},
\end{eqnarray}
where $G_{MN}$ is the Einstein's tensor, and $T_{MN}$ is the stress-energy tensor defined in the usual manner as
\begin{equation}\label{stress}
T_{MN}=-\frac{2}{\sqrt{-g}}\frac{\delta\left(\sqrt{-g}\mathcal L_m\right)}{\delta g^{MN}}.
\end{equation}

Two equations of motion describing the dynamics of the scalar fields $\varphi$ and $\psi$ can also be obtained as follows. We start with a variation of Eq. \eqref{act2} with respect to $\varphi$ and $\psi$. The resultant equations of motion will depend on the Ricci scalar $R$. Taking the trace of Eq. \eqref{eqfield}, we are able to cancel the terms depending on $R$ but we are left with a system of two coupled differential equations both depending on $\Box\varphi$ and $\Box\psi$. These two equations can be algebraically manipulated to isolate the terms $\Box\varphi$ and $\Box\psi$, so to isolate the dynamics of each scalar field in its own equation. The resultant equations of motion for $\varphi$ and $\psi$ are
\begin{equation}\label{eqphi}
\Box\varphi+\frac{1}{8}\left[5V-3\left(\varphi V_\varphi+\psi V_\psi\right)\right]=\frac{\kappa^2}{4}T,
\end{equation}
\begin{equation}\label{eqpsi}
\Box\psi-\frac{1}{2\psi}\partial^P\psi\partial_P\psi-\frac{3}{8}\psi\left(V_\psi+V_\varphi\right)=0,
\end{equation}
respectively, where the subscripted potentials $V_\varphi$ and $V_\psi$ denote derivatives of the potential $V\left(\varphi,\psi\right)$ with respect to the scalar fields $\varphi$ and $\psi$, respectively, and $T=g^{MN}T_{MN}$ is the trace of the stress-energy tensor.

%%%%%%%%%%%%%%%%%%%%%%%%%%%%%%%%%%%%%%%%%%%%%%%%%%%%%%%%%%%%%%%%%%%%
\section{Static and flat brane models with scalar field matter}\label{sec:models}
%%%%%%%%%%%%%%%%%%%%%%%%%%%%%%%%%%%%%%%%%%%%%%%%%%%%%%%%%%%%%%%%%%%%

In this section, we will consider matter to be described by a single dynamical scalar field $\chi$ with an associated interaction potential $U\left(\chi\right)$. The matter action describing this distribution of matter is thus
\begin{equation}\label{actmatter}
S_m=-\int_\Omega\sqrt{-g}\left[\frac{1}{2}\partial^P\chi\partial_P\chi+U\left(\chi\right)\right]d^5x.
\end{equation}
The stress-energy tensor $T_{MN}$ associated with this matter distribution can be computed via a variation of Eq. \eqref{actmatter} with respect to the scalar field $\chi$ and the definition of $T_{MN}$ provided in Eq. \eqref{stress}. We thus obtain
\begin{equation}\label{stresschi}
T_{MN}=-g_{MN}\left[\frac{1}{2}\partial^P\chi\partial_P\chi+U\left(\chi\right)\right]+\partial_M\chi\partial_N\chi.
\end{equation}
Furthermore, and similarly to the scalar fields $\varphi$ and $\psi$, an equation of motion for the field $\chi$ can be obtained varying Eq. \eqref{actmatter} with respect to $\chi$. As the field $\chi$ is minimally coupled to the metric $g_{MN}$, a dynamical equation for $\chi$ is immediately obtained with no need for further manipulations. The resultant equation is
\begin{equation}\label{eqchi}
\Box\chi=U_\chi,
\end{equation}
where the subscripted potential $U_\chi$ denotes a derivative of the potential $U$ with respect to the scalar field $\chi$. Our complete system of equations thus consists of Eqs. \eqref{eqfield}, \eqref{eqphi}, \eqref{eqpsi} and \eqref{eqchi}, with the stress-energy tensor $T_{MN}$ given by Eq. \eqref{stresschi}.

For the purpose of this paper, let us consider the static 5-dimensional line element 
\begin{equation}\label{metric}
ds^2=e^{2A\left(y\right)}\eta_{\mu\nu} dx^\mu dx^\nu + dy^2,
\end{equation}
where $A\left(y\right)$ is called the warp function, $\eta_{\mu\nu}$ is the 4-dimensional Minkowski metric given by $\eta_{\mu\nu}=\text{diag}\left(-1,1,1,1\right)$, and $y$ represents the extra 5th dimension of infinite extent. We shall assume that both the gravitational scalar fields $\varphi=\varphi\left(y\right)$ and $\psi=\psi\left(y\right)$ and the matter scalar field $\chi=\chi\left(y\right)$ are constant throughout the 4-dimensional spacetime and vary solely across the extra dimension $y$. Furthermore, given the isotropy of the 4-dimensional part of the metric, only two independent field equations arise. Inserting these assumptions and the metric from Eq.\eqref{metric} into the system of Eqs. \eqref{eqfield}, \eqref{eqphi}, \eqref{eqpsi} and \eqref{eqchi} yields
\begin{eqnarray}\label{field1}
3\left(2A'^2+A''\right)\left(\varphi-\psi\right)+3A'\left(\varphi'-\psi'\right)+\frac{2\psi'^2}{3\psi}
	\nonumber \\
+\frac{V}{2}+\varphi''-\psi''=-\frac{\kappa^2}{2}\left(\chi'^2+2U\right),
\end{eqnarray}
\begin{eqnarray}\label{field2}
6A'^2\left(\varphi-\psi\right)+4A'\left(\varphi'-\psi'\right)-\frac{2\psi'^2}{3\psi}
	\nonumber \\
+\frac{V}{2}=\frac{\kappa^2}{2}\left(\chi'^2-2 U\right),
\end{eqnarray}
for the two independent field equations,
\begin{eqnarray}\label{kgphi}
\varphi''+4A'\varphi'+\frac{1}{8}\left[5V-3\left(\varphi V_\varphi+\psi V_\psi\right)\right]
	\nonumber \\
=-\frac{\kappa^2}{8}\left(3\chi'^2+10 U\right),
\end{eqnarray}
\begin{equation}\label{kgpsi}
\psi''+4A'\psi'-\frac{\psi'^2}{2\psi}-\frac{3\psi}{8}\left(V_\psi+V_\varphi\right)=0,
\end{equation}
for the gravitational scalar fields $\varphi$ and $\psi$, respectively, and
\begin{equation}\label{kgchi}
\chi''+4A'\chi'-U_\chi=0,
\end{equation}
for the matter scalar field $\chi$. One can prove that Eq. \eqref{kgchi} is not an independent equation for the system in the following manner: take a derivative with respect to $y$ of Eq. \eqref{field2}, use Eq. \eqref{field1} to cancel the terms depending on $\zeta''$, then use Eqs. \eqref{kgphi} and \eqref{kgpsi} to cancel the terms depending on $\varphi''$ and $\psi''$ respectively, and finally use Eq. \eqref{field2} itself to cancel the term depending on $V\left(\varphi,\psi\right)$. The result of the stated algebraic manipulations is Eq. \eqref{kgchi}, thus proving its dependence on the remaining equations. Consequently, the system of Eqs. \eqref{field1} to \eqref{kgpsi} fully describes the system in study. The system consists of four independent equations for the six independent variables $A$, $\varphi$, $\psi$, $\chi$, $V$ and $W$.

%%%%%%%%%%%%%%%%%%%%%%%%%%%%%%%%%%%%%%%%%%%%%%%%%%%%%%%%%%%%%%%%%%%%
\subsection{The standard GR case}
%%%%%%%%%%%%%%%%%%%%%%%%%%%%%%%%%%%%%%%%%%%%%%%%%%%%%%%%%%%%%%%%%%%%

The well-known GR standard case can be obtained from the system of Eqs. \eqref{field1} to \eqref{kgchi}. This particular case corresponds to a form of the function $f\left(R,\mathcal R\right)$ as
\begin{equation}\label{grcase}
f\left(R,\mathcal R\right)=f\left(R\right)=R.
\end{equation} 
Using the definition of the scalar field provided in Eq. \eqref{scalars}, one verifies that $\varphi=1$ and $\psi=0$. Also, from Eq. \eqref{potential}, one verifies that in this case $V=0$. Thus, to obtain the standard case, one takes the limit $\varphi\to 1$, $\psi\to 0$, and $V\to0$ of the modified field equations in Eqs. \eqref{field1} and \eqref{field2}, which become respectively
\begin{equation}\label{field1gr}
A''=-\frac{\kappa^2}{3}\chi'^2,
\end{equation}
\begin{equation}\label{field2gr}
A'^2=\frac{\kappa^2}{12}\left(\chi'^2-2U\right),
\end{equation}
where we have used Eq. \eqref{field2gr} to cancel the dependency of Eq. \eqref{field1gr} in $A'$. On the other hand, the particular form of the function $f$ for the standard case given by Eq. \eqref{grcase} does not allow for an equivalent scalar-tensor representation of the theory due to the fact that the determinant of its Jacobian matrix vanishes. This implies that the relationship between the scalar fields $\varphi$ and $\psi$ with $R$ and $\mathcal R$ is degenerate. The equations of motion for the scalar fields, i.e., Eqs. \eqref{kgphi} and \eqref{kgpsi} are thus effectively removed from the system.

Furthermore, one can prove that the equation of motion for the scalar field $\chi$ given by Eq. \eqref{kgchi} is not independent of the two equations above. To do so, one takes the derivative of Eq. \eqref{field2gr} with respect to $y$ and uses Eq. \eqref{field1gr} to cancel the term depending on $A''$, obtaining Eq. \eqref{kgchi} as a result.

It is common to encounter Eqs. \eqref{field1gr} and \eqref{field2gr} in the literature with $\kappa^2=2$. To ease the comparison between our results in the upcoming sections and the available literature, we shall consider $\kappa^2=2$ from this point onward.

%%%%%%%%%%%%%%%%%%%%%%%%%%%%%%%%%%%%%%%%%%%%%%%%%%%%%%%%%%%%%%%%%%%%
\subsection{Solution without matter ($\chi=0$)}\label{sec:nomat}
%%%%%%%%%%%%%%%%%%%%%%%%%%%%%%%%%%%%%%%%%%%%%%%%%%%%%%%%%%%%%%%%%%%%

For simplicity, let us start by considering a brane model supported solely by the scalar fields $\varphi$ and $\psi$, i.e., without the matter field $\chi$, which results in Eq. \eqref{kgchi} being automatically satisfied. Furthermore, we shall assume that the dependency of the potential $V$ in the scalar fields $\varphi$ and $\psi$ is of the form $V\left(\varphi-\psi\right)$, as commonly considered in literature \cite{Rosa:2017jld,Rosa:2018jwp}. This choice is made so that the potential is a function of the coupling between $R$ and the scalar fields $\varphi$ and $\psi$ in Eq.\eqref{act2}. In this case, the partial derivatives of $V$ become related by $V_\varphi=-V_\psi\equiv \hat V$. Subtracting Eq. \eqref{field2} from Eq. \eqref{field1} and imposing the assumptions mentioned above leads to
\begin{equation}\label{fieldnomat}
3\left(\varphi-\psi\right)A''-\left(\varphi'-\psi'\right)A'+\frac{4\psi'^2}{3\psi}+\varphi''-\psi''=0.
\end{equation}
On the other hand, the equations for the scalar fields given in Eqs. \eqref{kgphi} and \eqref{kgpsi} become
\begin{equation}\label{kgphinomat}
\varphi''+4A'\varphi'+\frac{1}{8}\left[5V-3\left(\varphi-\psi\right)\hat V\right]=0,
\end{equation}
\begin{equation}\label{kgpsinomat}
\psi''+4A'\psi'-\frac{\psi'^2}{2\psi}=0.
\end{equation}
Finally, an equation relating $V$ to $\hat V$ arises from a derivative of $V\left(\varphi\left(y\right),\psi\left(y\right)\right)$ with respect to $y$ using the chain rule, which results in 
\begin{equation}\label{potnomat}
\frac{dV}{dy}=\hat V\left(y\right)\left(\varphi'-\psi'\right). 
\end{equation}

The system of Eqs. \eqref{fieldnomat} to \eqref{potnomat} consists of a system of four independent equations to the five unknowns $\varphi$, $\psi$, $V$, $\hat V$, and $A$. Thus, the system is under-determined and an extra constraint must be imposed to close the system. We choose to specify the form of the warp function A to be the usual form for a thick-brane solution:
\begin{equation}\label{warpnomat}
A\left(y\right)= A_0 \ln\left[\text{sech}\left(k y\right)\right],
\end{equation}
where $A_0$ and $k$ are constants and $A_0$ in particular must be defined positive. Let us focus on even solutions for the scalar fields $\varphi$ and $\psi$ (and consequently $V$). These solutions must satisfy the boundary conditions $\varphi'\left(0\right)=\psi'\left(0\right)=0$. Furthermore, let us denote $\varphi\left(0\right)=\varphi_0$ and $\psi\left(0\right)=\psi_0$ their values at the origin. In the same way, we will have $V'\left(0\right)=0$ and $V\left(0\right)=V_0$. Inserting these boundary conditions into Eq. \eqref{fieldnomat} yields
\begin{equation}\label{derivnomat}
\varphi''-\psi''=3A_0k^2\left(\varphi_0-\psi_0\right).
\end{equation}
In order to preserve the positivity of the factor $\varphi-\psi$ in Eq. \eqref{act2}, we should impose $\varphi_0>\psi_0$. This ensures that at $y=0$ we have $\varphi''\left(0\right)>\psi''\left(0\right)$. We shall look for solutions for which the positiveness of this factor is maintained throughout the entire range of $y$.

\begin{figure}
\includegraphics[scale=0.9]{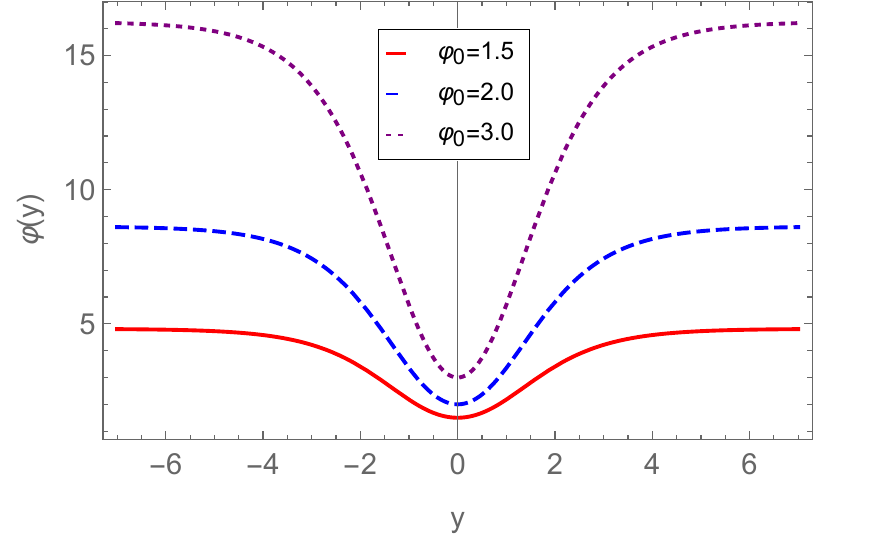}
\caption{Numerical solutions for the scalar field $\varphi\left(y\right)$ from Sec. \ref{sec:nomat} with $\psi_0=1$, $V_0=1$, $A_0=1$ and $k=1$. The scalar field satisfies the boundary condition $\varphi\left(0\right)=\varphi_0$ and approaches an asymptotically constant value for $|y|\gg 1$, as expected.}
\label{fig:phinomat}
\end{figure}

As can be seen from Eq. \eqref{kgpsinomat}, imposing a boundary condition $\psi'\left(0\right)=0$ will also set $\psi''\left(0\right)=0$, and thus the only possible even solution for $\psi$ is the constant solution $\psi=\psi_0$. We shall thus consider the constant $\psi_0$ as a free parameter and do not provide plots for the function $\psi\left(y\right)$. On the other hand, the numerical solutions for $\varphi\left(y\right)$ are plotted in Fig. \ref{fig:phinomat}. These solutions satisfy the boundary condition $\varphi\left(0\right)=0$, grow outwards from $y=0$ as expected from the positiveness of $\varphi''$ arising from Eq. \eqref{derivnomat} with $\psi''=0$, and approach a constant value for $|y|\gg 1$. The shape of these solutions is independent of the boundary conditions imposed for $V$ and $\psi$, but they should only be considered as long as $\varphi_0>\psi_0$.
%%%%%%%55555%5
%%%%%%%%%%%%%%%%

\begin{figure*}
\includegraphics[scale=0.9]{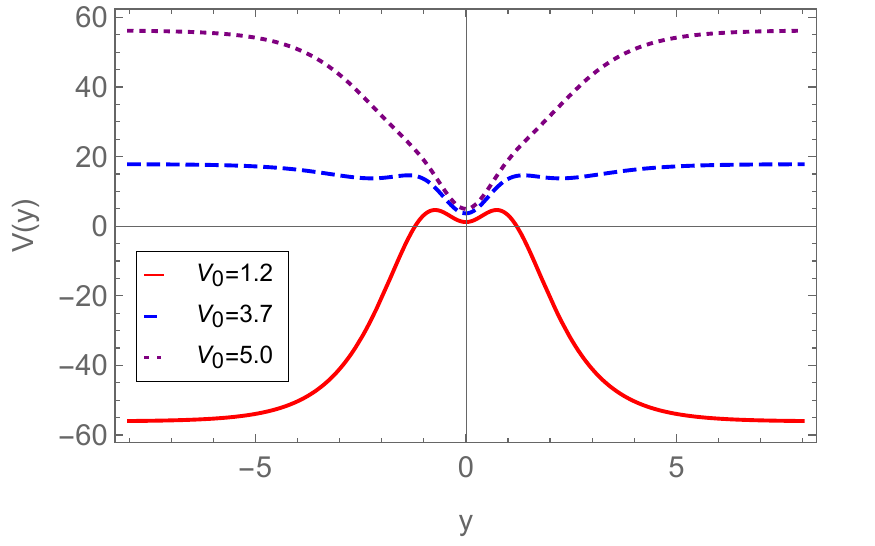}
\includegraphics[scale=0.9]{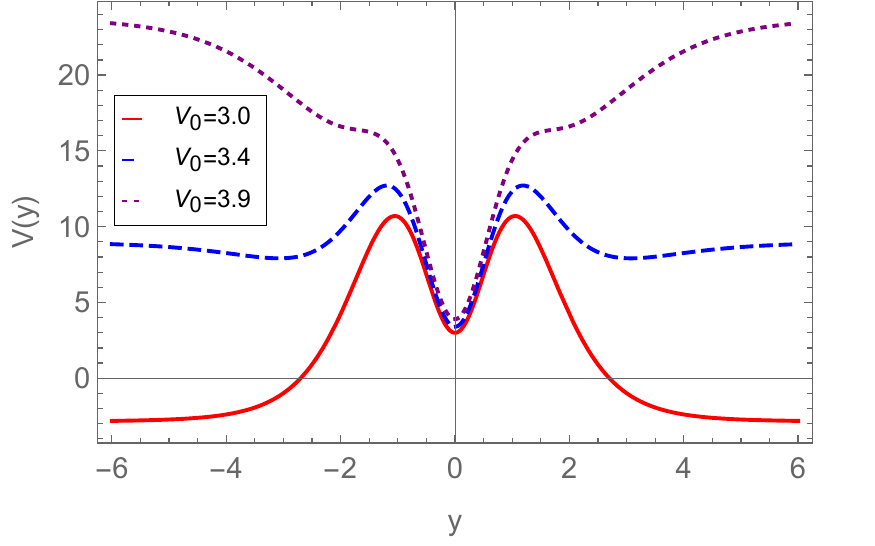}
\caption{Numerical solutions for the potential $V\left(y\right)$ from Sec. \ref{sec:nomat} with $\varphi_0=2$, $\psi_0=1$, $A_0=1$ and $k=1$. In the left panel we provide solutions for the boundary conditions $V_0=1.2$, $V_0=3.7$ and $V_0=5.0$. It is visible that there is a transition in the behavior of $V\left(y\right)$ depending on $V_0$. In the right panel, we provide the solutions with $V_0=3.0$, $V_0=3.4$ and $V_0=3.9$ for a more detailed view of this transition.}
\label{fig:potnomat}
\end{figure*}

Numerical solutions for the potential $V\left(y\right)$ are provided in Fig. \ref{fig:potnomat}. We see in the left panel that for $V_0\lesssim 1$ the potential has a positive concavity at the origin but eventually turns over and approaches asymptotically a negative constant. On the other hand, for $V_0\gtrsim 5$ the potential is always increasing and approaches asymptotically a positive constant. The transition between these two behaviors occurs at $V_0\sim 3.5$. A more detailed view of this transition is provided in the right panel.
%%%%%%%%%%%%%%%%%%%%%%%%%
%%%%%%%%%%%%%%%%%%%%%%%%%
The solutions provided in this section are consistent with the required localization of gravity; see Sec. \ref{sec:perturb} for further details.

%%%%%%%%%%%%%%%%%%%%%%%%%%%%%%%%%%%%%%%%%%%%%%%%%%%%%%%%%%%%%%%%%%%%
\subsection{Solution with matter 1: Ansatz for the matter field $\chi$ and the matter potential $U$}\label{sec:withmat}
%%%%%%%%%%%%%%%%%%%%%%%%%%%%%%%%%%%%%%%%%%%%%%%%%%%%%%%%%%%%%%%%%%%%

In this section, we shall obtain a solution for a thick brain in the presence of the matter field $\chi$. In order to close the system of Eqs. \eqref{field1} to \eqref{kgpsi} one has to impose two constraints into the system. We choose to set the forms of the potential $U\left(\chi\right)$ and the scalar field $\chi$ as
\begin{equation}\label{Uwithmat}
U\left(\chi\right)=\frac{1}{2}W_\chi^2-\frac{4}{3}W\left(\chi\right)^2,
\end{equation}
\begin{equation}\label{chiwithmat}
\chi\left(y\right)= \text{tanh}\ y,
\end{equation}
respectively, where $W\left(\chi\right)$ is called the super-potential of $\chi$ and takes the form
\begin{equation}\label{Wwithmat}
W\left(\chi\right)=\chi-\frac{1}{3}\chi^3.
\end{equation}
The motivation behind the forms of the mater field $\chi$, the potential $U$ and the super-potential $W$ selected in Eqs. \eqref{Uwithmat} to \eqref{Wwithmat} is their close connection to the standard GR case, which have motivated these forms to be widely used in the literature. Inserting Eqs. \eqref{Uwithmat} to \eqref{Wwithmat} into Eq. \eqref{kgchi} provides a differential equation for the warp function $A$ which can be solved immediately to yield a solution of the form
\begin{equation}
A\left(y\right)=A_0-\frac{4}{9}\ln\left(\text{cosh}\ y\right)+\frac{1}{9}\text{sech}^2\ y,
\end{equation}
where $A_0$ is a dimensionless constant which is usually set in such a way that the warp function has a value at the origin $A\left(0\right)=0$. For this purpose, we set $A_0=-1/9$.

Similarly to the previous case, we assume that the dependency of the potential $V$ in $\varphi$ and $\psi$ is of the form $V\left(\varphi-\psi\right)$ in such a way that its derivatives become $V_\varphi=-V_\psi=\hat V$. Subtracting Eq. \eqref{field2} from Eq. \eqref{field1} and imposing the assumptions mentioned above leads to
\begin{eqnarray}
\varphi''-\psi''+\frac{4\psi'}{3\psi}-2\ \text{sech}^4y\left(\phi-\psi-1\right)
	\nonumber \\
+\frac{2}{9}\left(\varphi'-\psi'\right)\text{tanh}\ y\left(2+\text{sech}^2y\right)=0.\label{fieldwithmat}
\end{eqnarray}
The equations for the scalar fields given in Eqs. \eqref{kgphi} and \eqref{kgpsi} become then
\begin{eqnarray}
&&\varphi''-\frac{24}{27}\varphi'\left(2+\text{sech}^2y\right)\text{tanh}\ y
+\frac{1}{8}\left[5V-3\hat V\left(\varphi-\psi\right)\right]
	\nonumber \\
&&\qquad \qquad =\frac{2}{27}\left(5\ \text{sech}^6y+42\ \text{sech}^4y-20\right),
\label{kgphiwithmat} 
\end{eqnarray}
and
\begin{equation}\label{kgpsiwithmat}
\psi''-\frac{8}{9}\left[2+\text{cosh}\left(2y\right)\right]\text{sech}^2y\ \text{tanh}\ y\ \psi'-\frac{\psi'^2}{2\psi}=0,
\end{equation}
respectively.
Note that the previous relationship between $V$ and $\hat V$ given in Eq. \eqref{potnomat} still hold, as in this section we have imposed the same constraint on $V$ and defined $\hat V$ in the same way. 

The set of Eqs. \eqref{fieldwithmat} to \eqref{kgpsiwithmat}, along with the relationship provided in Eq. \eqref{potnomat} constitute a system of four equations to the four unknowns $\varphi$, $\psi$, $V$ and $\hat V$, and thus the system is determined. Again, we focus on even solutions for $\varphi$ and $\psi$ (and consequently $V$). To do so, we impose the same boundary conditions at the origin, i.e., $\varphi'\left(0\right)=\psi'\left(0\right)=0$, $\varphi\left(0\right)=\varphi_0$, and $\psi\left(0\right)=\psi_0$. As a consequence, we have also $V'\left(0\right)=0$. Inserting these boundary conditions into Eq. \eqref{fieldwithmat} leads to
\begin{equation}
\varphi''-\psi''=2\left(\varphi_0-\psi_0-1\right).
\end{equation}
As we have already mentioned, in order to preserve the positivity of the factor $\varphi-\psi$ we should impose $\varphi_0>\psi_0$.  However, to ensure that $\varphi''\left(0\right)>\psi''\left(0\right)$, we must also guarantee that $\varphi-\psi>1$, thus preserving the positiveness of the factor $\varphi-\psi$ throughout the whole range of $y$.

\begin{figure}
\includegraphics[scale=0.9]{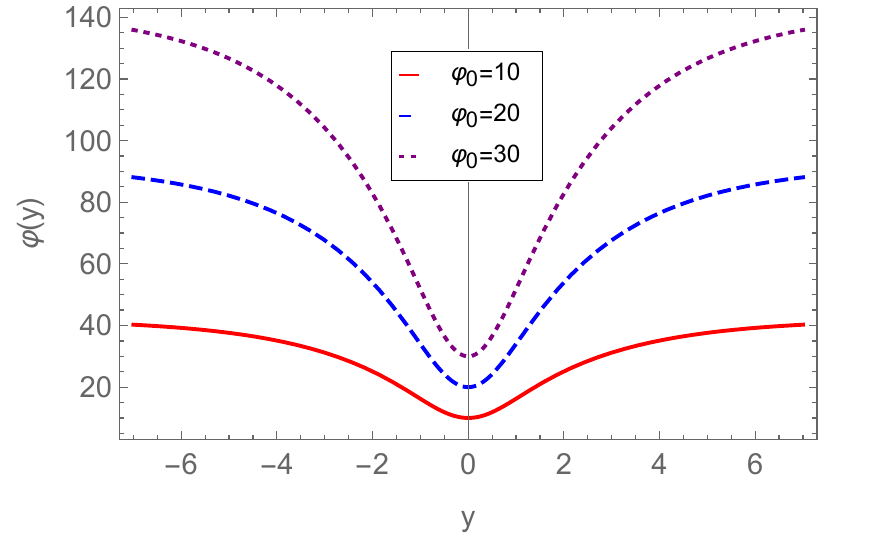}
\caption{Numerical solutions for the scalar field $\varphi\left(y\right)$ from Sec. \ref{sec:withmat} with $\psi_0=1$ and $V_0=1$. The scalar field satisfies the boundary condition $\varphi\left(0\right)=\varphi_0$ and approaches an asymptotically constant value for $|y|\gg 1$, as expected.}
\label{fig:phimat}
\end{figure}

\begin{figure}
\includegraphics[scale=0.9]{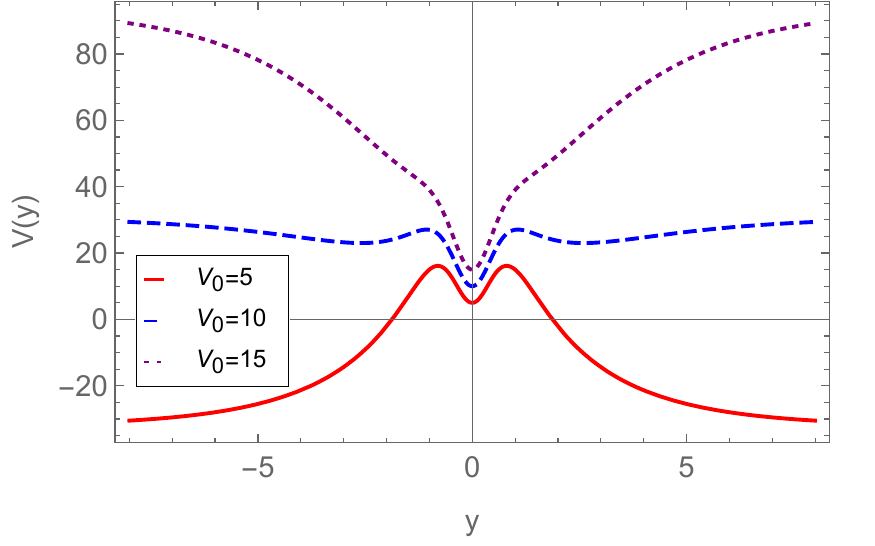}
\caption{Numerical solutions for the potential $V\left(y\right)$ from Sec. \ref{sec:withmat} with $\varphi_0=10$, and $\psi_0=1$. Similarly to the case without matter, the behavior of the potential is strongly dependent on the boundary conditions $V_0$ and $\varphi_0$.}
\label{fig:potmat}
\end{figure}

Similarly to the case without matter, imposing a boundary condition $\psi'\left(0\right)=0$ in Eq. \eqref{kgpsiwithmat} will consequently set $\psi''\left(0\right)=0$, and thus the only possible even solution for $\psi$ is the constant solution $\psi=\psi_0$. Therefore, $\psi_0$ takes again the role of a free parameter in the problem. The numerical solutions for the scalar field $\varphi$ are plotted in Fig. \ref{fig:phimat}. These solutions satisfy the boundary condition $\varphi'\left(0\right)=0$ and grow outwards from the origin, approaching a constant asymptotic value for $|y|\gg 1$. The shape of these solutions is independent of the boundary conditions for $V$ and $\psi$ but should only be considered for $\varphi_0-\psi_0>1$.
%%%%%%%%%%%%%%%%%%%%%%%%
%%%%%%%%%%%%%%%%%%
Finally, numerical solutions for the potential $V\left(y\right)$ are given in Fig. \ref{fig:potmat}. Similarly to the case without matter, the potential always presents a positive concavity at the origin but, depending on the boundary conditions $V_0$ and $\varphi_0$, its behavior might eventually turn over and approach negative values (for $V_0\lesssim 5$ with $\varphi_0=10$) or grow throughout the whole range of $y$ (for $V_0\gtrsim 15$ with $\varphi_0=10$). The turning point between the two behaviors occurs at roughly $V_0\sim 10$.
%%%%%%%%%%%%%%%%%%%%%
%%%%%%%%%%%%%%%%%%
The solutions provided in this section are consistent with the required localization of gravity, as we further describe in Sec. \ref{sec:perturb}.

%%%%%%%%%%%%%%%%%%%%%%%%%%%%%%%%%%%%%%%%%%%%%%%%%%%%%%%%
\subsection{Solution with matter 2: Ansatz for the warp function $A$ and the scalar field $\varphi$}\label{sec:withmat3}
%%%%%%%%%%%%%%%%%%%%%%%%%%%%%%%%%%%%%%%%%%%%%%%%%%%%%%%%

To complete the analysis, in this section we will derive another solution for a thick brane in the presence of a matter field $\chi$, thus still having the freedom to impose two constraints to close the system of Eqs. \eqref{field1} to \eqref{kgpsi}. In this case, we shall leave both the matter field $\chi$ and its potential $U$ as free functions and provide the following ansatze for the warp function $A$ and the scalar field $\varphi$:
\begin{equation}\label{warpmatfinal}
A\left(y\right)=A_0 \ln\left[\text{sech}\left(ky\right)\right],
\end{equation}
\begin{equation}\label{varphimatfinal}
\varphi\left(y\right)=\varphi_0\ \text{tanh}^2\left(ky\right),
\end{equation}
where $A_0$ and $\varphi_0$ are constants defined positive and $k$ is an arbitrary constant.

Following the same reasoning as in previous cases, we assume that the dependency of the potential $V$ in $\varphi$ and $\psi$ is of the form $V\left(\varphi-\psi\right)$ in such a way that its derivatives become $V_\varphi=-V_\psi=\hat V$. Subtracting Eq. \eqref{field2} from Eq. \eqref{field1} and inserting Eqs. \eqref{warpmatfinal} and \eqref{varphimatfinal} yields
\begin{eqnarray}
&&\psi''+A_0k\ \text{tanh}\left(ky\right)\psi'-\frac{4\psi'^2}{3\psi}
	\nonumber \\
&&\qquad -3A_0k^2\ \text{sech}^2\left(ky\right)\psi  =2\chi'^2  -\varphi_0k^2\text{sech}^2 \times
	\nonumber \\
&& \qquad \qquad \times \left(ky\right)\left[4+A_0-\left(6+A_0\right)\text{sech}^2\left(ky\right)\right].\label{fieldfinal}
\end{eqnarray}

Inserting the same Eqs. \eqref{warpmatfinal} and \eqref{varphimatfinal} into the Eqs. \eqref{kgphi} to \eqref{kgchi} yields the equations for the scalar fields
\begin{eqnarray}
&&\frac{1}{8}\left[5V-3\hat V\left(\varphi_0\ \text{tanh}^2\left(ky\right)-\psi\right)\right]+\frac{5}{2}U+\frac{3}{4}\chi'^2
	\nonumber \\
&& \qquad = 16\varphi_0k^2\left[\left(1\!+\!2A_0\right)\text{cosh}\left(2ky\right)\!-\!2\left(1\!+\!A_0\right)\right] \times
	\nonumber \\
&& \qquad  \qquad \qquad \times  \text{sech}^4\left(ky\right)\!,\label{kgphifinal}
\end{eqnarray}
\begin{equation}\label{kgpsifinal}
\psi''-4A_0k\ \text{tanh}\left(ky\right)\psi'-\frac{\psi'^2}{2\psi}=0,
\end{equation}
\begin{equation}\label{kgchifinal}
\chi''-4A_0k\ \text{tanh}\left(ky\right)\chi'+U_\chi=0,
\end{equation}
respectively.
Note that the previous relationship between $V$ and $\hat V$ given in Eq. \eqref{potnomat} still hold, as in this section we have imposed the same constraint on $V$ and defined $\hat V$ in the same way. The system of Eqs. \eqref{fieldfinal} to \eqref{kgchifinal}, along with Eq. \eqref{potnomat} is a system of five equations to the five unknowns $U$, $V$, $\hat V$, $\psi$, and $\chi$, and thus it is determined. A complete analytic solution for this system can be found through the following reasoning. We will not write explicitly the forms of the solutions due to their lengthy character. Instead we will provide the respective plots.

\begin{figure}
\includegraphics[scale=0.9]{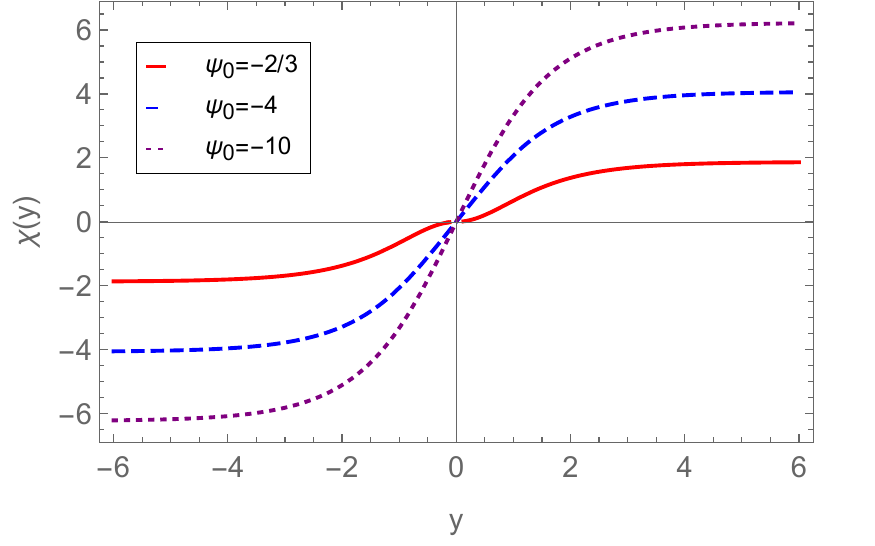}
\caption{Solutions for the matter field $\chi\left(y\right)$ from Sec. \ref{sec:withmat3} with $\varphi_0=1$, $A_0=1$ and $k=1$. The solutions cross the origin $y=0$ and approach asymptotically constant values for $|y|\gg 1$.}
\label{fig:chifinal}
\end{figure}

Focusing on even solutions for the scalar field $\psi$, we have to impose the boundary condition $\psi\left(0\right)=0$ into Eq. \eqref{kgpsifinal}. As a consequence, we obtain as before that $\psi''\left(0\right)=0$ and thus the only possible even solution for $\psi$ is the constant solution $\psi=\psi_0$, that we take to be a free parameter of the system. Now, inserting this result into Eq. \eqref{fieldfinal} yields a decoupled differential equation for $\chi$ which can be solved immediatly. The solutions of this equation will be real functions if the parameters satisfy the following relation:
\begin{equation}\label{psicrit}
\psi_0<-\frac{2\varphi_0}{3A_0}=\psi_c,
\end{equation}
where $\psi_c$ is the critical value. The solution obtained for $\chi$ is analytic and can be written in terms of elliptic functions. We provide a plot of this solution for different combinations of parameters in Fig. \ref{fig:chifinal}. For values of $\psi_0$ close to the critical value $\psi_c$ given in Eq. \eqref{psicrit}, the matter field $\chi$ has a triple step shape, which becomes less evident with a decrease in $\psi_0$.

%%%%%%%%%%%%%%
%%%%%%%%%%%%%%%

\begin{figure}
\includegraphics[scale=0.9]{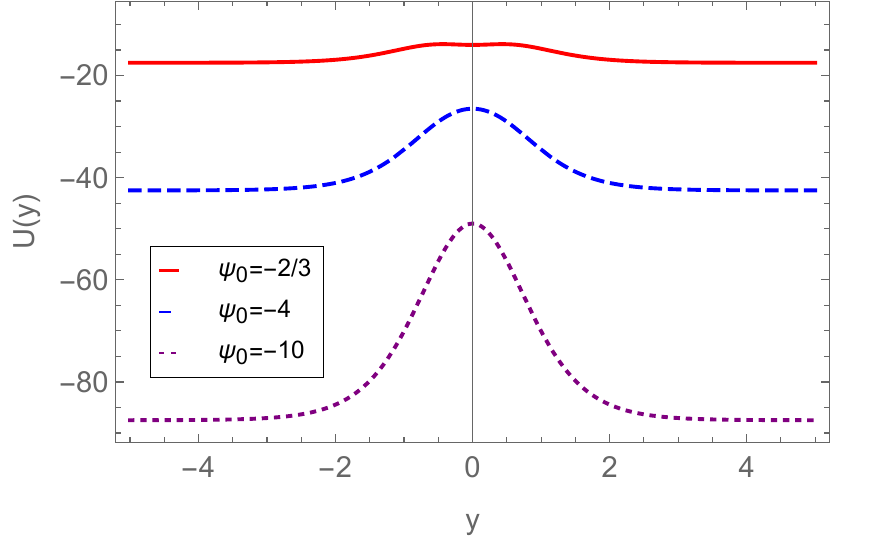}
\caption{Solutions for the matter potential $U\left(y\right)$ from Sec. \ref{sec:withmat3} with $\varphi_0=1$, $A_0=1$ and $k=1$. The potential gets steeper with a decrease in $\psi_0$.}
\label{fig:Uchifinal}
\end{figure}

Inserting the previous solution for $\chi$ into Eq. \eqref{kgchifinal} allows us to solve for $U'\left(\chi\right)$, which can then be transformed into $U'\left(y\right)$ via the chain rule and integrated to obtain a solutions for $U\left(y\right)$ as
\begin{eqnarray}
U\left(y\right)&=&\frac{k^2}{4}\text{sech}^2\left(ky\right)\left[\left(1+4A_0\right)\left(4\varphi_0+A_0\varphi_0-3A_0\psi_0\right)\right.
	\nonumber \\
&&-\left.\varphi_0\left(6+A_0\right)\left(1+2A_0\right)\text{sech}^2\left(ky\right)\right]\,.
\end{eqnarray}
We plot this solution for different combinations of parameters in Fig. \ref{fig:Uchifinal}. The shape of the potential does not change dramatically, but it gets steeper with a decrease in $\psi_0$.

%%%%%%%%%%%%%%%%%%%
%%%%%%%%%%%%%%%%%%%

Finally, Eq. \eqref{fieldfinal} can be solved with respect to $\hat V$ and the corresponding solution can be inserted in Eq. \eqref{potnomat} and directly integrated to obtain solutions for the potential $V$, which are
\begin{eqnarray}
V\left(y\right)&=&-2A_0k^2\left[6A_0\left(\varphi_0-\psi_0\right)
	\right.\nonumber \\
&&\left.+\varphi_0\left(2-5A_0\text{cosh}\left(2ky\right)\right)\text{sech}^4\left(ky\right)\right]\,.
\end{eqnarray}

\begin{figure}
\includegraphics[scale=0.9]{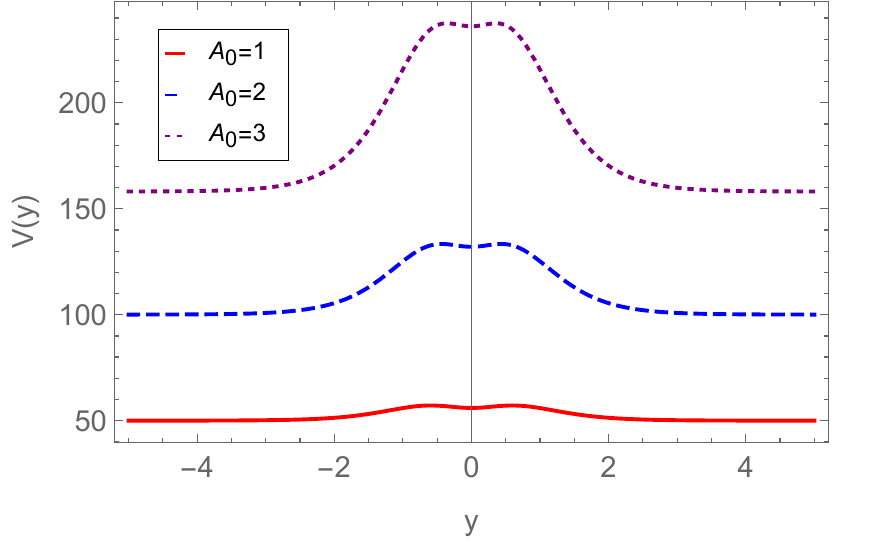}
\caption{Solutions for the potential $V\left(y\right)$ from Sec. \ref{sec:withmat3} with $\varphi_0=1$, $\psi_0=-\frac{2}{3}$ and $k=1$. Unlike previous cases, the general shape of the potential is always the same for the allowed range of parameters.}
\label{fig:potmat3}
\end{figure}

These solutions are plotted in Fig. \ref{fig:potmat3}. Unlike the previous cases studied where a change in the parameters of the problem could change the shape of the potential drastically, in this case the general shape of the potential remains the same because we are restricted by the inequality in Eq. \eqref{psicrit}.

%%%%%%%%%%%%%%%%%%%%
%%%%%%%%%%%%%%%%%%%

The solutions provided in this section are consistent with the required localization of gravity; see Sec. \ref{sec:perturb} for further details.

%%%%%%%%%%%%%%%%%%%%%%%%%%%%%%%%%%%%%%%%%%%%%%%%%%%%%%%%%%%%%%%%%%%%
\section{Metric perturbations}\label{sec:perturb}
%%%%%%%%%%%%%%%%%%%%%%%%%%%%%%%%%%%%%%%%%%%%%%%%%%%%%%%%%%%%%%%%%%%%

%%%%%%%%%%%%%%%%%%%%%%%%
\begin{figure*}
\includegraphics[scale=0.9]{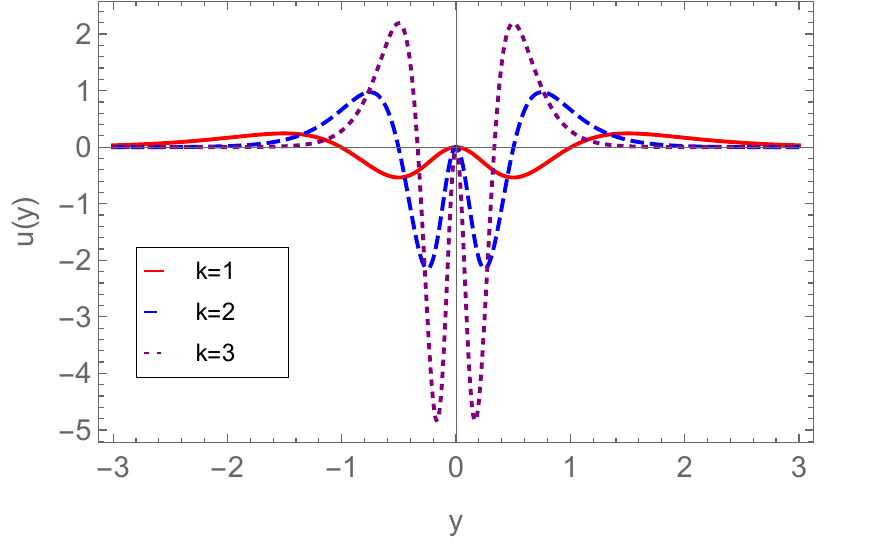}
\includegraphics[scale=0.9]{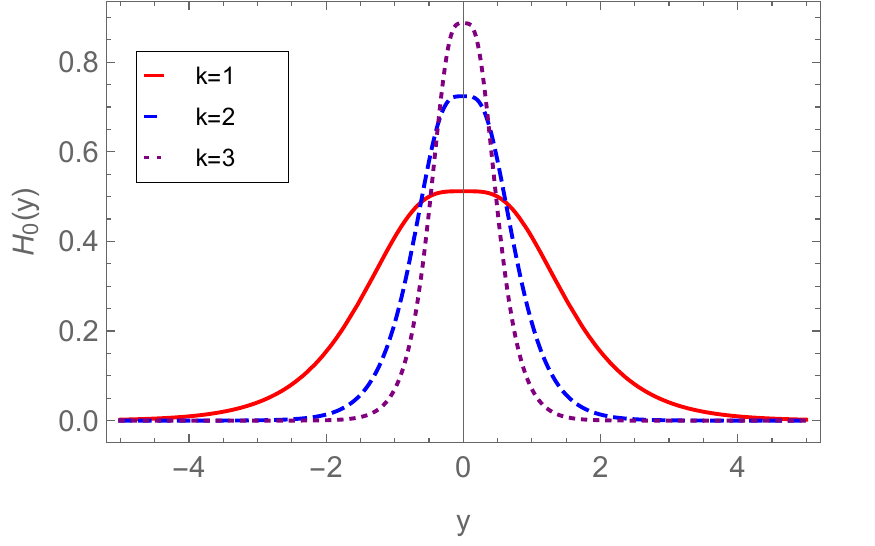}
\caption{Potential $u\left(y\right)$ (left panel) and graviton zero mode $H_0\left(y\right)$ (right panel) for the solutions described in Sec. \ref{sec:nomat} with $\varphi_0=2$, $\psi_0=1$, $A_0=1$ and $V_0=1$. The shape of the potential $U$ is always a double well. The finiteness of the integral of $H_0$ implies that the zero graviton mode is localized on the brane.}
\label{fig:normnomat}
\end{figure*}
%%%%%%%%%%%%%%%%%%%

%%%%%%%%%%%%%%%%%%%%%%%%
\begin{figure*}
\includegraphics[scale=0.9]{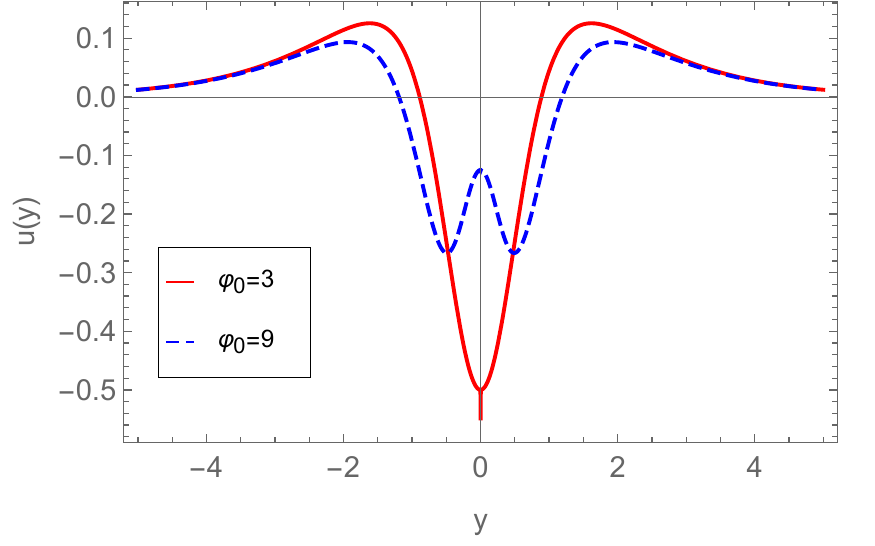}
\includegraphics[scale=0.9]{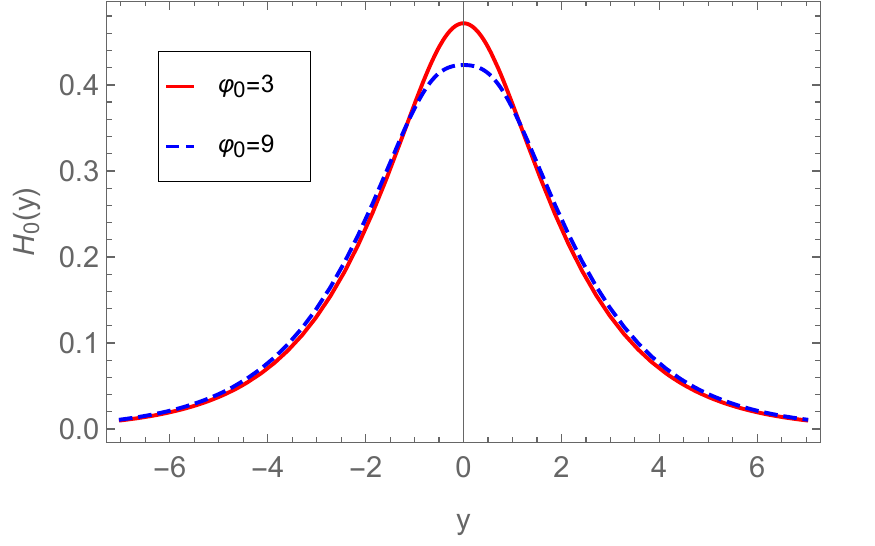}
\caption{Potential $u\left(y\right)$ (left panel) and graviton zero mode $H_0\left(y\right)$ (right panel) for the solutions described in Sec. \ref{sec:withmat} with $V_0=1$ and $\psi_0=1$. Depending on the choice of $\varphi_0$, the potential $U$ might be a single or a double well. The finiteness of the integral of $H_0$ implies that the zero graviton mode is localized.}
\label{fig:normmat}
\end{figure*}
%%%%%%%%%%%%%%%%%%%%%%

In this section, we will consider small perturbations of the metric in the form
\begin{equation}\label{metric2}
ds^2=e^{2A\left(y\right)}\left(\eta_{\mu\nu}+H_{\mu\nu}(x,y)\right) dx^\mu dx^\nu + dy^2.
\end{equation}
In addition, we will assume small perturbations on the scalar field $\chi$ such that $\chi \rightarrow \chi + \tilde{\chi}(x,y)$. In this case, one can shown that the ($\mu\nu$) components of the linearized field equations are
\begin{eqnarray}\label{field2p}
&&(\varphi-\psi)\bigg\{-e^{2A}\left[\frac{1}{2}H''_{\mu\nu}+2A'H'_{\mu\nu}+\frac{1}{2}\left(\ln (\varphi-\psi)\right)' H'_{\mu\nu}\right]
	\nonumber \\
&&-\frac{1}{2}\square^{(4)}H_{\mu\nu}-\frac{1}{2}\eta^{\alpha\beta}\left(\partial_{\mu}\partial_{\nu}H_{\alpha\beta}-\partial_{\mu}\partial_{\alpha}H_{\nu\beta}-\partial_{\nu}\partial_{\alpha}H_{\mu\beta}\right)
\nonumber \\
&& \qquad \qquad +\frac{1}{2}e^{2A}\eta_{\mu\nu}H'\left[\left(\ln (\varphi-\psi)\right)'-A'\right]\bigg\}
	\nonumber \\
&& \qquad \qquad \qquad =\frac{2\kappa^2}{3}e^{2A}\eta_{\mu\nu}U_{\chi}\tilde{\chi} \,,
\end{eqnarray}
where $\square^{(4)}=\eta^{\mu\nu}\partial_{\mu}\partial_{\nu}$. By using the transverse and traceless gauge ($\partial^{\mu}H_{\mu\nu}=0$ and $\eta^{\mu\nu}H_{\mu\nu}=H=0$), the metric perturbation decouples from the scalar and the above equation reduces to
\begin{equation}\label{linearized}
H''_{\mu\nu}+\left[4A'+\left(\ln (\varphi-\psi)\right)'\right] H'_{\mu\nu}+e^{-2A}\square^{(4)}H_{\mu\nu}=0.
\end{equation}

In order to better understand how this equation relates to stability, we will transform it into a Schr\"odinger-like equation. First, we make a coordinate transformation $dy=e^{A}dz$ which makes the metric \eqref{metric} conformally flat. In the new variable $z$, the Eq. \eqref{linearized} can be written as
\begin{equation}\label{graviton}
\partial_{z}^{2}H_{\mu\nu}+\left(3\partial_{z}A+\partial_{z}\ln (\varphi-\psi)\right)\partial_{z}H_{\mu\nu}+\square^{(4)}H_{\mu\nu}=0.
\end{equation}
Now, we remove the term with first order derivative in Eq. \eqref{graviton} by the following redefinition of the tensor perturbation:
\begin{equation}
H_{\mu\nu}(x,z)=\frac{e^{-3A(z)/2}}{\sqrt{\varphi(z)-\psi(z)}}\hat{H}_{\mu\nu}(x,z).
\end{equation}
Thus, Eq. \eqref{graviton} takes the form
\begin{equation}
-\partial_{z}^{2}\hat{H}_{\mu\nu}+\left(\alpha^{2}(z)-\partial_{z}\alpha(z)\right)\hat{H}_{\mu\nu}-\square^{(4)}\hat{H}_{\mu\nu}=0,
\end{equation}
where we have defined 
\begin{equation}
\alpha(z) \equiv -\frac{3}{2}\partial_{z}A-\frac{1}{2}\partial_{z}\ln(\varphi-\psi).
\end{equation}

Finally, we perform the decomposition $\hat{H}_{\mu\nu}(x,z)=\xi_{\mu\nu}(x)H(z)$ to obtain the Klein-Gordon equation $\square^{(4)}\xi_{\mu\nu}(x)=m^{2}\xi_{\mu\nu}(x)$ and the Schr\"odinger-like equation
\begin{equation}\label{Schro}
\left[-\partial_{z}^{2}+u(z)\right]H(z)=m^{2}H(z),
\end{equation}
where
\begin{equation}\label{potbarrier}
u(z)=\alpha^{2}-\partial_{z}\alpha.
\end{equation}
Note that in Eq. \eqref{Schro} the zero mode ($m^{2}=0$) represents the massless graviton, while the Kaluza-Klein modes ($m^{2}>0$) represent massive excitations. Note also that Eq. \eqref{Schro} can be factorized as
\begin{equation}
Q^{\dagger}QH(z)=m^{2}H(z),
\end{equation}
with operators $Q$ and $Q^{\dagger}$ given by
\begin{equation}
Q=\partial_{z}+\alpha(z)\quad \text{and} \qquad Q^{\dagger}=-\partial_{z}+\alpha(z).
\end{equation}
This factorization shows that the Schr\"odinger-like equation cannot support states with negative eigenvalues, i.e., $m^{2}\geq0$. Thus, the system is stable against small perturbations of the metric.

The zero mode is obtained by performing $QH_{0}(z)=0$. The result is
\begin{equation}
H_{0}(z)={\cal N}\sqrt{\varphi-\psi}\,\,e^{3A/2},
\end{equation}
where $\cal{N}$ is a normalization constant. To ensure that 4-dimensional gravity can be recovered on the brane, the zero mode must be normalizable. The normalization constant can thus be obtained from
\begin{equation}\label{normalize}
\int H_0^2dz=\mathcal N^2\int (\varphi-\psi)e^{2A}dy=1.
\end{equation}

For the model without matter from Sec. \ref{sec:nomat}, it can be show that the integral in Eq. \eqref{normalize} is finite and the normalization constant $\mathcal N$ can be computed for different combinations of parameters. It can also be shown that the shape of the potential $u\left(y\right)$ given in Eq. \eqref{potbarrier} is independent of the boundary conditions considered for $\varphi$, $\psi$ and $V$, being only affected by changes in the warp function $A$. 
Thus, in Fig. \ref{fig:normnomat} we display both the potential $u\left(y\right)$ and the zero modes $H_0\left(y\right)$ for different values of $k$, keeping the remaining free parameters constant. Similarly to what happens in the non-generalized version of the hybrid metric-Palatini gravity, the shape of the potential $u$ is always a double well. Given the finiteness of the integral of the zero mode $H_0$, the graviton zero mode can thus be localized on the brane.

\begin{figure*}
\includegraphics[scale=0.9]{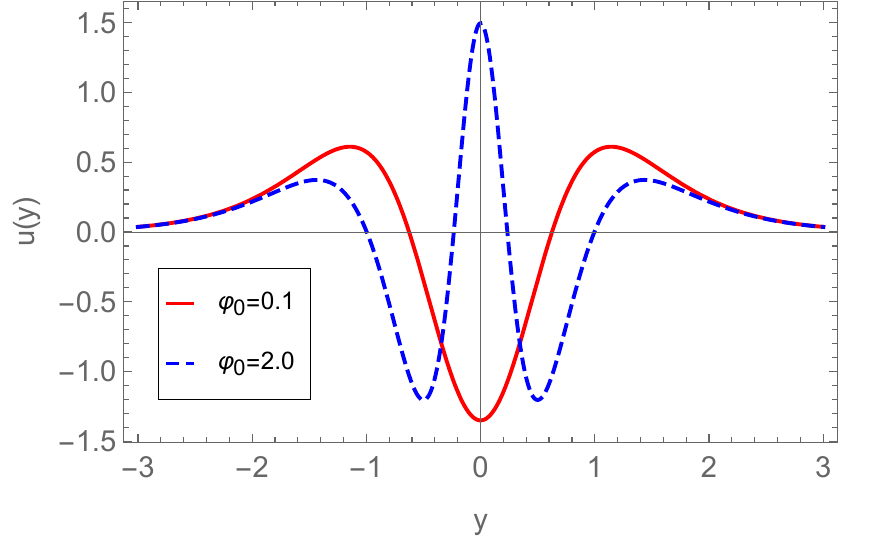}
\includegraphics[scale=0.9]{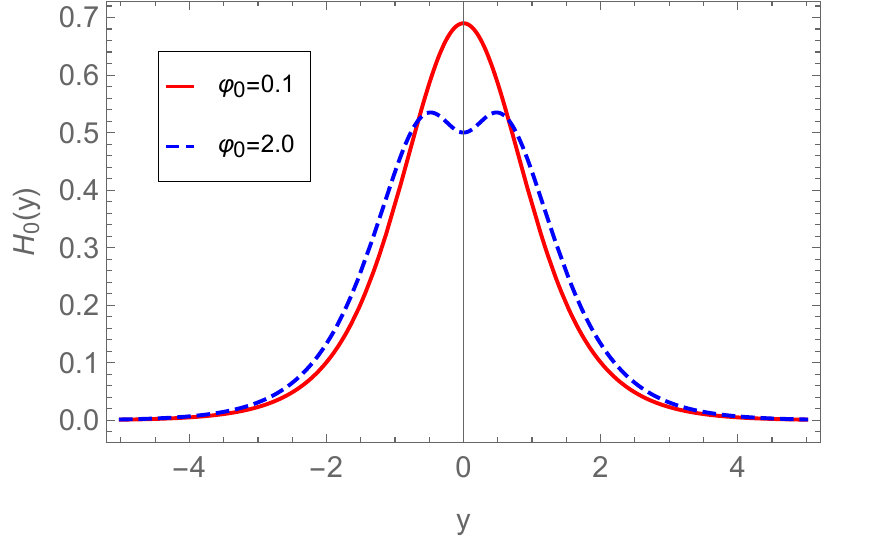}
\caption{Potential $u\left(y\right)$ (left panel) and graviton zero mode $H_0\left(y\right)$ (right panel) for the solutions described in Sec. \ref{sec:withmat3} with $A_0=1$, $\psi_0=-\frac{2}{3}$, and $k=1$. Depending on the choice of $\varphi_0$, the potential $U$ might be a single or a double well, potentially leading to the graviton mode to be split in two peaks. The finiteness of the integral of $H_0$ implies that the zero graviton mode is localized.}
\label{fig:normmat3}
\end{figure*}

For the models featuring a matter field $\chi$ in Sec. \ref{sec:withmat} and  Sec. \ref{sec:withmat3}, the integral in Eq. \eqref{normalize} is again finite and the normalization constant $\mathcal N$ can be computed for different combinations of parameters. The main difference between the case without matter and the teo studied cases in the presence of a matter field $\chi$ is that the shape of potential $u\left(y\right)$ can be either a single or a double well depending on the choice of parameters, unlike the case without matter where the potential was always a double well, see Figs. \ref{fig:normmat}, and \ref{fig:normmat3}. For the model in Sec. \ref{sec:withmat}, a change of the global shape of the potential $u$ does not imply a dramatic change in the graviton zero mode. However, for the model in Sec. \ref{sec:withmat3}, under an appropriate choice of parameters, this change of shape in the potential implies that the graviton zero mode $H_0$ is localized in a wider region around the origin, suggesting that the braneworld seems to support internal structure.  Given the finiteness of the integral of $H_0$ over $y$, one concludes that for both the cases with matter the generalized hybrid metric-Palatini braneworld is stable against fluctuations in the metric.

%%%%%%%%%%%%%%%%%%%%
%%%%%%%%%%%%%%%%%%%%%%

%%%%%%%%%%%%%%%%%%%%%%%%%%%%%%%%%%%%%%%%%%%%%%%%%%%%%%%%%%%%%%%%%%%%
\section{Summary and Discussion}\label{sec:concl}
%%%%%%%%%%%%%%%%%%%%%%%%%%%%%%%%%%%%%%%%%%%%%%%%%%%%%%%

As is well-known, in the standard braneworld scenario, the presence of braneworld configurations with internal structure appeared, in the presence of two scalar fields, in the form of a Bloch brane \cite{Bazeia:2004dh}. This suggests that one could investigate braneworld scenarios within the scalar-tensor representation of the generalized hybrid-Palatini gravity, which consists of two distinct scalar fields, and was in fact explored in the present work. Indeed, an interesting  possibility would be to consider two extra fields with standard dynamics, as considered in \cite{Bazeia:2004dh}. Another possibility would be to consider the two extra fields with modified dynamics, as recently considered in \cite{douglas}. The third model investigated in Ref. \cite{douglas}, in particular, was shown to engender an interesting mechanism to induce internal structure into the brane, so it can also be used in the present context to show its efficiency concerning the induction of internal structure in the brane. 

More specifically, in this work, we studied several 5-dimensional braneworld scenarios within the context of the generalized hybrid metric-Palatini gravity. We investigated two distinct cases, namely, the presence and the absence of the extra scalar field, $\chi$. In the absence of the field $\chi$, we found an interesting solution, and in the presence of $\chi$, we studied two distinct possibilities. One of the models in the presence of a matter field $\chi$ showed that variations of parameters do not add new qualitative effects in the profile of the braneworld solutions. However, the third case studied in Sec. \ref{sec:withmat3} showed an interesting effect which was also found in \cite{Fu:2016szo}, concerning the spreading of the zero mode. Although this new effect is different, it suggests the possibility to study braneworld configurations that engender internal structure. 

The presence of internal structure has also appeared before, in the context of modified theories of gravity with non-constant curvature \cite{B1,B3,correia}, so this can also be considered in the present context of generalized metric-Palatini gravity. Moreover, is would be of interest to understand how the mechanism used in \cite{ahmed,douglas} to make the brane asymmetric can be extended to the novel scenario described in the present work. Another issue of current interest concerns the entrapment of fermions and gauge fields inside the brane within the generalized metric-Palatini gravity context explored in the present work. The above issues deserve further investigations, and we are now considering some possibilities, hoping to report on them in the near future.

%%%%%%%%%%%%%%%%%%%%%%%%%%%%%%%%%%%%%%%%%%%%%%%%%%%%%%%%%%%%%%%%%%%%

%%%%%%%%%%%%%%%%%%%%%%%%%%%%%%%%%%%%%%%%%%%%%%%%%%%%%%%%%%%%%%%%%%%%
\begin{acknowledgments}
JLR was supported by the European Regional Development Fund and the programme Mobilitas Pluss (MOBJD647). DB is supported by Conselho Nacional de Desenvolvimento Cient\'\i fico e Tecnol\'ogico, grants No. 404913/2018-0 and No. 303469/2019-6, and by Paraiba State Research Foundation, grant No. 0015/2019. FSNL acknowledges support from the Funda\c{c}\~{a}o para a Ci\^{e}ncia e a Tecnologia (FCT) Scientific Employment Stimulus contract with reference CEECIND/04057/2017, and the FCT research grants No. UID/FIS/04434/2020, No. PTDC/FIS-OUT/29048/2017 and No. CERN/FIS-PAR/0037/2019.
\end{acknowledgments}
%%%%%%%%%%%%%%%%%%%%%%%%%%%%%%%%%%%%%%%%%%%%%%%%%%%%%%%%%%%%%%%%%%%%

%%%%%%%%%%%%%%%%%%%%%%%%%%%%%%%%%%%%%%%%%%%%%%%%%%%%%%%%%%%%%%%%%%%%

%%%%%%%%%%%%%%%%%%%%%%%%%%%%%%%%%%%%%%%%%%%%%%%%%%%%%%%%%%%%%%%%%%%%

%%%%%%%%%%%%%%%%%%%%%%%%%%%%%%%%%%%%%%%%%%%%%%%%%%%%%%%%%%%%%%%%%%%%
\end{document}